\documentclass[12pt]{article}
\usepackage{graphicx}
\newcommand\beq{\begin{equation}}
\newcommand\bear{\begin{eqnarray}}
\newcommand\eeq{\end{equation}}
\newcommand\eear{\end{eqnarray}}
\begin{document}
\baselineskip=24pt

\begin{center}
{\Large \bf Charge-transfer induced large nonlinear optical properties 
of small Al Clusters: Al$_{4}$M$_{4}$ (M=Li, Na and K)} 
\end{center}

\vspace*{0.5cm}

\centerline{\bf Ayan Datta and Swapan K. Pati$^*$}

\vspace*{0.1cm}

\begin{center}
Theoretical Sciences Unit and Chemistry and Physics of Materials Unit \\
Jawaharlal Nehru Center for Advanced Scientific Research  \\
Jakkur Campus, Bangalore 560 064, India. E-mail: pati@jncasr.ac.in
\end{center}

\begin{center}
{\bf Abstract}
\end{center}

\vspace*{0.2cm}

We investigate the linear and nonlinear electric polarizabilities of small
Al$_{4}$M$_{4}$ (M=Li, Na and K) clusters. Quantum chemical calculations reveal
that these compounds exhibit an exceptionally high magnitude of linear and
nonlinear optical (NLO) coefficients which are orders of magnitude
higher than the conventional $\pi$-conjugated systems of similar sizes.
We attribute such phenomenal increase to non-centrosymmetricity
incorporated in the systems by the alkali atoms surrounding the
ring leading to charge transfer with small optical gap and low bond length
alternation (BLA). Such a low magnitude of the BLA from a different origin,
 suggests the possibility that these clusters are aromatic in character and
along with the large NLO coefficients they appear to be better candidates
for next generation NLO fabrication devices.

\newpage
\clearpage

The development of materials with large nonlinear optical (NLO) properties
is a key to controlling the propagation of light by optical means. In
particular, the response of the materials to the application of the
electric field has found tremendous applications in designing materials
for NLO devices\cite{nlo1}. These devices are being used in numerous
applications, from lasers to optical switches and optoelectronics.
The NLO properties of organic $\pi$-conjugated materials
have been studied in great details in the last few decades 
\cite{wills,marks}. 
The second and third order non-linear optical properties, $\beta$ and 
$\gamma$ for the $\pi$-conjugated polymers increase with the 
conjugation length (L) roughly as $L^3$ and $L^5$ respectively
\cite{pugh}. Therefore, the general strategy to model NLO materials
has been to increase the conjugation length. However, there exist
an upper limit for every off-resonant susceptibilities 
\cite{Kuzyk}. Alternatives to these
$\pi$-conjugated compounds are yet to be explored theoretically 
in a detailed fashion. But, with the gaining popularity of various 
{\it ab-initio} level methods\cite{Helgaker}, there has been a 
tremendous impetus in 
investigating the structure and electronic properties of both 
homogeneous and heterogeneous small clusters in recent years\cite{c1,c2}.
 
Small Al$_{4}$ rings like Al$_{4}$M$_{4}$ and their anions Al$_{4}$M$_{3}^{-}$, 
M=alkali metals, have been a subject of
current interest \cite{sharan,aroma} because of their unique characteristics 
and close structural resemblance with the C$_{4}$H$_{4}$. 
However, although C$_{4}$H$_{4}$ is an anti-aromatic species, these
Al$_{4}$-clusters are recently reported to be $\sigma$ aromatic\cite{schleyer}. 
Thus, it would be interesting to ask whether these rings are better 
polarized 
than their organic counterpart; whether the structural characteristics has 
any role in their polarization response functions. Organic $\pi$-conjugated 
systems are stabilized due to $\pi$-electron delocalizations, while 
the inorganic metal complexes reduce their energy through strong charge 
transfer. There have been no previous efforts to study in details the NLO 
properties of these all-metal clusters. 
We describe in the following that these metal clusters offer a unique 
polarization response due to their ionic character, contrary to conventional 
$\pi$-conjugated systems, leading to large optical coefficients.

We begin our calculations by optimizing the ground state geometries of 
the Al$_{4}$-clusters (Al$_{4}$Li$_{4}$, Al$_{4}$Na$_{4}$ and Al$_{4}$K$_{4}$). All the 
optimizations have been done using the 6-31G(d,p) basis set. 
Electron correlation has been included according to the DFT method 
using Becke's three parameter hybrid formalism 
and the Lee-Yang-Parr functionals (B3LYP)\cite{dft} available 
in the GAMESS\cite{gamess} electronic structure set of codes.
Since, we want to compare the optical properties of these small four-membered
rings with their organic analogue C$_{4}$H$_{4}$, we start with a planar  
initial geometry for the optimizations. We have varied the level of basis 
set from 6-31G(d,p) to 6-311G+(d) to ensure that these geometries 
correspond to the minima in the potential energy surface. The final 
geometries indeed remain independent of the selection of the
basis set. Contrary to that of C$_{4}$H$_{4}$ having
a rectangular ring, these Al$_{4}$-clusters are found to have a rhomboidal 
structure with four Al atoms forming a rhombus and the four alkali atoms
around the four Al-Al bonds  forming four Al-M (Li, Na, K)-Al triangles. One 
of the diagonals of the Al$_{4}$ ring are also connected by Al-Al bond. 
The equilibrium geometries are shown in Fig. 1 [1(a), 1(b) and 1(c)].

While Al$_{4}$Li$_{4}$ and Al$_{4}$K$_{4}$ have a planar structure (D$_{2h}$),
Al$_{4}$Na$_{4}$ has a distorted structure, with the four Na atoms arranged in 
a nonplanar geometry around the planar ring (the Na atoms are distorted by 13 
degrees from the plane of Al$_{4}$ ring). This can be understood by considering
the increase in size of the alkali ions and the distances of the ions from 
the Al$_4$-ring. With the progressive increase in the ionic radii of 
counterion, Li 
to K(Li=0.68 $\AA$, Na=0.97 $\AA$ and K = 1.33 $\AA$), the structures are expected 
to be distorted and the four alkali atoms should arrange in a non-centrosymmetric
geometry around the Al$_{4}$ ring to minimize steric repulsion. But, the 
average Al-M distance increases while going from Al$_{4}$Li$_{4}$ (2.65 $\AA$)
to Al$_{4}$Na$_{4}$ (3.00 $\AA$) to Al$_{4}$K$_{4}$ (3.35 $\AA$). Although 
the ionic radii of K ion is more than that of Li and Na, in Al$_{4}$K$_{4}$,  
the four K ions are far separated from the Al$_{4}$ ring, allowing a planar 
structure. For Al$_{4}$Na$_{4}$, both the ionic radius of Na and the average 
Al-M distance fall in between Al$_{4}$Li$_{4}$ and 
Al$_{4}$K$_{4}$ and thereby minimizes the steric repulsion through distortion. 
   
Also, very close in energy to these planar rhomboidal structures for these
Al$_{4}$-clusters are the capped octahedron structures for 
the Al$_{4}$Li$_{4}$ [2(a)] and Al$_{4}$Na$_{4}$ [2(b)] 
(with C$_{2h}$ symmetry) and a distorted tricycle-like structure 
for Al$_{4}$K$_{4}$ [2(c)]. These geometries are shown in Fig.2. 
At the footnote of each structure, the corresponding ground state energies 
are given. It has however not escaped our attention that previous 
works on alkali derivatives of Al$_{4}$-clusters have predicted more 
than one unique structures for these systems\cite{kanhare}. This calls for a
study to elucidate whether the optical 
properties for these Al$_{4}$-clusters for different geometries are substantially 
different or are very similar. Hence, both the geometries for each cluster were 
considered for computing the optical response functions. 

These geometries were used to compute the SCF MO energies and then the
spectroscopic properties using the Zerner's INDO method\cite{zerner}.
We have varied the levels of CI calculations, with singles(SCI) and
multi-reference doubles CI (MRDCI), to obtain a reliable estimate of the
second order optical response. The later method is particularly important
since it includes correlation effects substantially. The MRDCI approach
adopted here has been extensively used in earlier works, and was found
to provide excitation energies and dipole matrix elements in good
agreement with experiment\cite{mrdci1,mrdci2}. As reference
determinants, we have chosen those
determinants which are dominant in the description of the ground state
and the lowest one-photon excited states\cite{mrdci3}. We report the
MRDCI results with 4 reference determinants including the 
Hartree-Fock HF ground state. For each
reference determinant, we use 5 occupied and 5 unoccupied molecular
orbitals to construct a CI space with configuration dimension of $800$ to
$900$. To calculate NLO properties, we use correction vector method, which
implicitly assumes all the excitations to be approximated by a correction
vector\cite{sr1}. Given the Hamiltonian matrix, the ground state wave
function and the dipole matrix, all in CI basis, it is straightforward to
compute the dynamic nonlinear optic coefficients using either the first
order or the second order correction vectors. Details of this method
have been published in a number of papers\cite{sr2,pati1,pati2}.

Table 1. shows the bond-length alternation (BLA), $\Delta r$, the
optical gap and the average Mulliken charge on the Al$_{4}$ ring for all
the geometries. The $\Delta r$ is defined as the average difference between
the bond lengths of two consecutive bonds in the $Al_4$-ring and the optical
gap is calculated as the energy difference between the geometry relaxed ground
state and the lowest optically allowed state with substantial oscillator
strength. This corresponds to the vertical absorption gap. 
To directly compare the efficiency of these Al$_{4}$-clusters 
with the conventional $\pi$ conjugated systems, we calculate the 
optical properties of the 1,3-cyclobutadiene (C$_{4}$H$_{4}$) and benzene 
(C$_{6}$H$_{6}$) at the same level of theory. For the Al$_{4}$-clusters, 
there is a substantial amount of charge transfer from the alkali atoms
to the Al atoms (negative charge), making them act as donor and acceptor
respectively. Such a charge transfer induces polarization in the
ground state structure and reduces the optical gap.
On the other hand, the C-H bond being perfectly covalent,
there is almost no charge transfer in case of C$_{4}$H$_{4}$ and C$_{6}$H$_{6}$ 
and thus have a large optical gap due to finite size molecular architecture. 

Charge transfer stabilizes the system with very small changes in the bond lengths. 
The chemical hardness, $\eta$, defined as, one half of (ionization potential-electron 
affinity), decreases
as one moves from Li to K. More specifically, the $\eta$ for Li, Na and K 
are $2.39$eV, $2.30$eV and $1.92$eV respectively\cite{pearson}. 
So, the extent of charge transfer from the alkali atom to the Al$_{4}$-ring 
should increase with the decrease in the chemical hardness of the alkali atoms 
which is evident from Table 1. From Al$_{4}$Li$_{4}$ to Al$_{4}$Na$_{4}$ 
to Al$_{4}$K$_{4}$, 
the Mulliken charge on the Al$_{4}$-ring increases leading to decrease in the 
BLA along the series with the exception of Al$_{4}$Na$_{4}$ [1(b) and 2(b)] 
which has much lower Mulliken charge on the Al$_{4}$-ring. 
For, the Al$_{4}$Na$_{4}$ [1(b)] as mentioned above, there is a substantial 
distortion of the Na atoms from the Al$_{4}$-ring. For the Al$_{4}$Na$_{4}$,
 with C$_{2h}$ symmetry [2(b)] even though there is no distortion, the large 
Al-Na distance reduces the ionicity of the bond. 
As a result the extent of charge transfer is lesser for Al$_{4}$Na$_{4}$.     

Benzene is aromatic with $\Delta r=0$. Although BLA can not be regarded 
as the sole measuring index of aromatic character, the small BLA 
found for the Al$_{4}$-clusters (together with large BLA 
for the anti-aromatic C$_4$H$_4$) tend to suggest that the 
Al$_{4}$-clusters are 
more like aromatic but most certainly not anti-aromatic species, as has been 
proposed recently\cite{aroma}. The distorted Al$_{4}$K$_{4}$ structure [2(c)] 
is very interesting. The Al$_{4}$-ring is distorted from planarity 
by $9.5$ degrees. Such a distortion arises to minimize the steric repulsion 
in accommodating four bulky K atoms on a plane, very similar to that of
cyclooctatetraene, which undergoes a distortion 
from planar to tub-shaped geometry to minimize the ring-strain\cite{nonaroma}.
Thus, this structure of Al$_{4}$K$_{4}$ is neither aromatic nor 
anti-aromatic but can be considered to be non-aromatic just like 
cyclooctatetraene. This is supported by the energies for the structures 
for Al$_{4}$K$_{4}$ [1(c) and 2(c)]. The distorted structure is more 
stable than the planar structure. Thus, the steric repulsion for the 
four large K atoms overwhelms the stability of the planar undistorted 
aromatic Al$_{4}$K$_{4}$ making the Al$_{4}$K$_{4}$ cluster non-aromatic.

In Table 2, the magnitudes of the ground state dipole moment, $\mu_G$, the 
linear($\alpha$), and nonlinear ($\beta$ and $\gamma$) 
polarizabilities for the clusters are reported from the ZINDO 
calculations. Note that we report the magnitudes for the tumbling 
averaged $\bar \alpha$, $\bar \beta$ and $\bar \gamma$, defined as\cite{tumbling}
\begin{eqnarray}
\bar \alpha & = & \frac{1}{3}\sum_{i}(\alpha_{ii}) \nonumber \\
\bar \beta & = & \sqrt{\sum_i \beta_i\beta_i^*};~~~~ \beta_i = \frac{1}{3}\sum_{j}{(\beta_{ijj}+\beta_{jij}+\beta_{jji})} \nonumber \\
\bar \gamma & = & \frac{1}{15}\sum_{ij}(2\gamma_{iijj}+\gamma_{ijji}) 
\end{eqnarray}
\noindent where the sums are over the coordinates $x,y,z$ ($i,j=x,y,z$) and 
$\beta_i^*$ refers to the conjugate of $\beta_i$ vector.
 
The ground state dipole moment $\mu_G$ and $\bar\beta$ are zero for the
C$_{4}$H$_{4}$ and C$_{6}$H$_{6}$ due to its perfect centrosymmetric geometry, 
although, the $\bar\alpha$ and $\bar\gamma$ have finite values. For the 
Al$_{4}$-clusters with the progressive increase in the ionic radii of counterion, 
the ground state
dipole moment increases. Thus, while Al$_{4}$Li$_{4}$ has no ground state 
dipole moment, Al$_{4}$Na$_{4}$ and Al$_{4}$K$_{4}$ have substantial ground 
state dipole moment (particularly 1(b) and 2(c) due to their non-centrosymmetric
 structures discussed above). For Al$_{4}$K$_{4}$, while the rhomboidal 
geometry has a very low ground state dipole moment but 
the distorted tricycle-like structure has a very high dipole moment.
Thus, due to the out-of-plane charge transfer, the dipole matrix
elements are also larger, resulting in particularly large value for 
$\bar\beta$. For Al$_{4}$Li$_{4}$ and 
Al$_{4}$Na$_{4}$ [2(b)] the polarization is in the excited state as the ground 
state dipole moment is zero. However, it is not the case for the insulating 
C$_{4}$H$_{4}$ and C$_{6}$H$_{6}$ which have zero polarization both in the 
ground and the optical excited states. Thus, $\bar\beta$ is zero for C$_{4}$H$_{4}$ 
and C$_{6}$H$_{6}$.

The optically active states are the low-energy 
states of these metallic clusters and the lowest optical gap is about 0.07 au 
for Al$_{4}$-clusters
compared to 0.25 au for the C$_{4}$H$_{4}$ and C$_{6}$H$_{6}$. 
Since the optical coefficients are inversely proportional to the optical gaps 
and proportional to the dipolar matrices, a large optical gap implies low 
magnitudes for the optical coefficients. C$_{4}$H$_{4}$ has the highest 
magnitude of BLA and optical gap and the least charge transfer on the 
ring structure, thereby smallest magnitude of $\bar\gamma$. On the otherhand, 
although BLA is zero for C$_{6}$H$_{6}$ due to complete $\pi$-electron 
delocalization, there is no charge transfer in the finite molecular structure 
leading to large optical gap and weak polarization. Consequently, $\bar\gamma$ 
is very less also for C$_{6}$H$_{6}$. 

In contrast, the optical coefficients in general are quite large for the 
Al$_{4}$-clusters. For example, the $\bar\gamma$ for the Al$_{4}$-clusters are 
roughly $10^{4}$ times more than that for C$_{4}$H$_{4}$ and C$_{6}$H$_{6}$. 
This is because the $\bar\gamma$ 
is a third order property with 4 dipolar matrices in the numerator and 3 
optical gaps in the denominator\cite{rat}. The $\bar\gamma$ for the 
Al$_{4}$-clusters 
increases with the increase in the polarization of the Al-M bonds and 
follows the trend: $\bar\gamma$ of Al$_4$Li$_4$ $<$ $\bar\gamma$ of 
Al$_4$Na$_4$ $<$ $\bar\gamma$ of  Al$_4$K$_4$ (same trend as $\eta$). 
But, the distorted 
structures for Al$_{4}$Na$_{4}$ and Al$_{4}$K$_{4}$ [1(b) and 2(c)]
have less $\bar\gamma$ due to less polarization of the Al-M bonds. 
For C$_{4}$H$_{4}$ 
($\Delta r$= 0.245$\AA$), C$_{6}$H$_{6}$ ($\Delta r$= 0.00$\AA$) and 
linear chain, (-CH=CH-)$_n$, n=3 ($\Delta r$= 0.1$\AA$) the $\bar\gamma$ 
are 2.21, 2.63 and 192.85 (all in au) per CH bond, respectively. This 
is in agreement with previous findings that the magnitude of $\gamma$ 
varies nonlinearly with $\Delta r$ and its maxima occurs at an optimal 
$\Delta r$ $\ne 0$ \cite{marder}. However, for the Al$_{4}$-clusters, 
because of strong charge-transfer, the $\bar\gamma$ are less sensitive 
to variation in $\Delta r$ and even for $\Delta r$=0.0$\AA$ 
(perfect square Al$_{4}$ ring) and $\Delta r$=0.245$\AA$ 
(rectangular Al$_4$ ring like C$_{4}$H$_{4}$) the $\bar\gamma$ 
are similarly high, as found for the optimized structures. Such 
charge-transfer induced large NLO is evident
for even bulk materials like CsLiB$_6$O$_{10}$ (CLBO) showing
 significant 4$^{th}$ and 5$^{th}$ harmonic generation \cite{CLBO}. 

The maximum possible value for the off-resonance $\gamma$(Al$_{4}$M$_{4}$)
calculated using the Kuzyk's simple two-state model \cite{Kuzyk} is about 2825
times more than that for $\gamma$(C$_{4}$H$_{4}$). This is in very good
agreement with our MRDCI calculations at a low frequency (0.001 au), 
which predict $\gamma$(Al$_{4}$M$_{4}$)/$\gamma$(C$_{4}$H$_{4}$) 
$\approx$ $10^4$.

To compare and contrast these clusters with their organic counterparts,
we calculate the NLO properties of the well-known $\pi$-conjugated systems, 
the {\it trans}-polyacetylene chain, (-CH=CH-)$_n$, by varying the number 
of spacers, $n$, from n=1 to 6, and thereby extending the length
of conjugation from 2.65 au to 29.11 au. The geometries were optimized by 
the same method as mentioned above. The linear and nonlinear 
polarizations are calculated at the same frequency ($0.001$ au). 
Our calculated values for the optical properties compare fairly well in 
trends with the experimental results that the linear ($\alpha$) and 
nonlinear ($\gamma$) optical properties increase steadily with the 
increase in the conjugation-length of the chain (see Table 2 and Table 3).
For example, for ethylene, $\gamma_{expt}$= 1504.9 au, for butadiene (n=2),
$\gamma_{expt}$= 4566.4 au and for hexatriene (n=3), $\gamma_{expt}$=
14950.1 au \cite{expt}. Note that our calculations are done at a
lower frequency compared to the laser frequency used in the experiment
(0.066 au). However, the magnitudes of all the polarization 
quantities are much higher for the charge-transfer complex 
(Al$_{4}$M$_{4}$ clusters) compared to the conventional $\pi$ conjugated 
chains with comparable conjugation length. 
Only when there are very large number of spacers (n=5-6), that the 
magnitudes become comparable to the much smaller Al$_{4}$-clusters. 
For example, $\gamma$[(-CH=CH-)$_6$] $\approx$ gamma(Al$_{4}$M$_{4}$). 
$(-CH=CH-)_6$ has 12 atoms in conjugation while Al$_{4}$M$_{4}$ has only 4.
So, as a thumb-rule, one can state that gamma for the Al-atoms in the
charge-transfer ring scale three times that for pi-conjugated organic
materials.
   
As discussed, the large NLO properties for Al$_{4}$M$_{4}$ is due 
to the charge transfer 
from the alkali metals to the Al$_{4}$ ring. It will be thus of interest 
to compare these hetero atomic all-metal clusters with alkylated organic 
compounds such as lithiated benzene or organolithium and organosodium derivatives like 
C$_{8}$H$_{6}$Li$_{2}$ and C$_{8}$H$_{6}$Na$_{2}$. These alkylated 
organic compounds also exhibit larger NLO coefficients. For example, 
$\gamma$(C$_{6}$Li$_{6}$)/$\gamma$(C$_{6}$H$_{6}$)=$7.3\times 10^{2}$ 
\cite{papa1}. 
Similarly, $\gamma$(C$_{8}$H$_{6}$Li$_{2}$)/$\gamma$(C$_{8}$H$_{6}$)=5.5 and 
$\gamma$(C$_{8}$H$_{6}$Na$_{2}$)/$\gamma$(C$_{8}$H$_{6}$)=20 \cite{papa2}. But, 
NLO responses for the Al$_{4}$M$_{4}$ is much higher than these 
alkylated organic compounds compared to pure organic materials. Also, 
there have been previous efforts to calculate the NLO coefficients 
in inorganic clusters like GaN, GaP and GaAs \cite{karna}. These systems 
have higher gap than the Al$_{4}$M$_{4}$ clusters so the NLO coefficients are 
smaller. 

To conclude, our theoretical study shows that the small four membered 
Al$_{4}$-clusters functionalized with various metal cations provide an
innovative route for selection of materials with very high nonlinear 
optical properties. We believe that our study will motivate experiments 
on these Al-clusters. Some of these compounds have already been well 
characterized from stable alloys\cite{aroma} but laser evaporation is not 
sufficient to stabilize these materials for NLO experiment. One idea 
will be to stabilize these clusters in a sandwitch type geometry by 
incorporating a suitable transition metal ion\cite{mercero} or solvents. 

SKP thanks CSIR and DST, Govt. of India, for the research grants.

\pagebreak
\clearpage

\newpage

\begin{table}
\caption{The bond length alternation, $\Delta r$ (in $\AA$),
Optical Gap (in au) and the average Mulliken charge ($\Delta q$) on the
ring for the clusters from ZINDO calculations.}
\end{table}

\begin{center}
\begin{tabular}{|l|l|l|l|}
\hline
Molecule  & $\Delta r$ & Gap & $\Delta q$     \\ \hline
Al$_{4}$Li$_{4}$ 1(a) & 0.1283 &  0.0819   & -0.592         \\ \hline
Al$_{4}$Li$_{4}$ 2(a) & 0.1276 &  0.024   & -0.506         \\ \hline
Al$_{4}$Na$_{4}$ 1(b) & 0.1302 &  0.0909 & -0.174           \\ \hline
Al$_{4}$Na$_{4}$ 2(b) & 0.1103 &  0.0607 & -0.127           \\ \hline
Al$_{4}$K$_{4}$  1(c) & 0.0656 & 0.0663   & -0.634         \\ \hline
Al$_{4}$K$_{4}$  2(c) & 0.0649 & 0.0867   & -0.618  \\ \hline
C$_{4}$H$_{4}$   & 0.245 &     0.2410    &  -0.030           \\ \hline
C$_{6}$H$_{6}$   & 0.000 &     0.2588    &  -0.009          \\ \hline

\end{tabular}
\end{center}

\pagebreak
\clearpage

\newpage

\begin{table}
\caption{The ground state dipole moment, $\mu_G$, linear polarizability, $\alpha$, 
1st hyperpolarizability, $\beta$ and the 2nd hyperpolarizability, $\gamma$, 
(tumbling average) for the clusters and for {\it trans}-
polyacetylene chain from ZINDO-MRDCI calculations. 
The units are in au. 'n' is the number of -$CH=CH$- units.}
\end{table}

\begin{center}
\begin{tabular}{|l|l|l|l|l|}
\hline
Molecule  & $\mu_G$ & $\bar \alpha$ & $\bar \beta$ & $\bar \gamma$  \\ \hline
Al$_{4}$Li$_{4}$ 1(a) & 0.000 & $4.9\times 10^{3}$   & 542.5  & $1.91\times 10^{7}$ \\ \hline
Al$_{4}$Li$_{4}$ 2(a) & 0.000 & $5.5\times 10^{3}$   & 244.9  & $5.33\times 10^{8}$ \\ \hline
Al$_{4}$Na$_{4}$ 1(b) & 0.076 & $5.9\times 10^{3}$   & 8465.2  & $1.09\times 10^{7}$ \\ \hline
Al$_{4}$Na$_{4}$ 2(b) & $8.6\times 10^{-4}$ & $8.7\times 10^{3}$   & 1098.5 & $2.00\times 1
0^{8}$    \\ \hline
Al$_{4}$K$_{4}$  1(c) & 0.004 & $5.4\times 10^{3}$   & 79.3 & $2.60\times 10^{7}$ \\ \hline
Al$_{4}$K$_{4}$ 2(c) & 5.720 & $4.7\times 10^{3}$  & $1.2\times 10^{5}$  & $1.90\times 10^{
7}$     \\ \hline
C$_4$H$_4$   & 0.000 & $2.9\times 10^{2}$   & 0.000  & $4.76\times 10^{3}$  \\ \hline
C$_{6}$H$_{6}$   & 0.000 & $5.4\times 10^{2}$    & 0.000  & $8.44\times 10^{3}$ \\ \hline
(CH=CH)$_n$, n=1  & 0.000 & 136.3 & 0.000 & $2.78\times 10^{4}$ \\ \hline
(CH=CH)$_n$, n=2  & 0.000 & 421.0 & 0.000  & $4.15\times 10^{4}$ \\ \hline
(CH=CH)$_n$, n=3  & 0.000 & 852.4 & 0.000  & $6.17\times 10^{5}$  \\ \hline
(CH=CH)$_n$, n=4  & 0.000 & 1455.2 & 0.000  & $2.82\times 10^{6}$ \\ \hline
(CH=CH)$_n$, n=5  & 0.000 & 2203.2  & 0.000 & $8.41\times 10^{6}$  \\ \hline
(CH=CH)$_n$, n=6  & 0.000 & 3074.9 & 0.000 & $2.07\times 10^{7}$  \\ \hline
\end{tabular}
\end{center}

\newpage
\clearpage

\begin{figure}
\caption{Equilibrium ground state geometries for Al$_{4}$Li$_{4}$,
Al$_{4}$Na$_{4}$ and Al$_{4}$K$_{4}$. The footnote
of each structure contains the ground state energies in au.}
\includegraphics[scale=0.8] {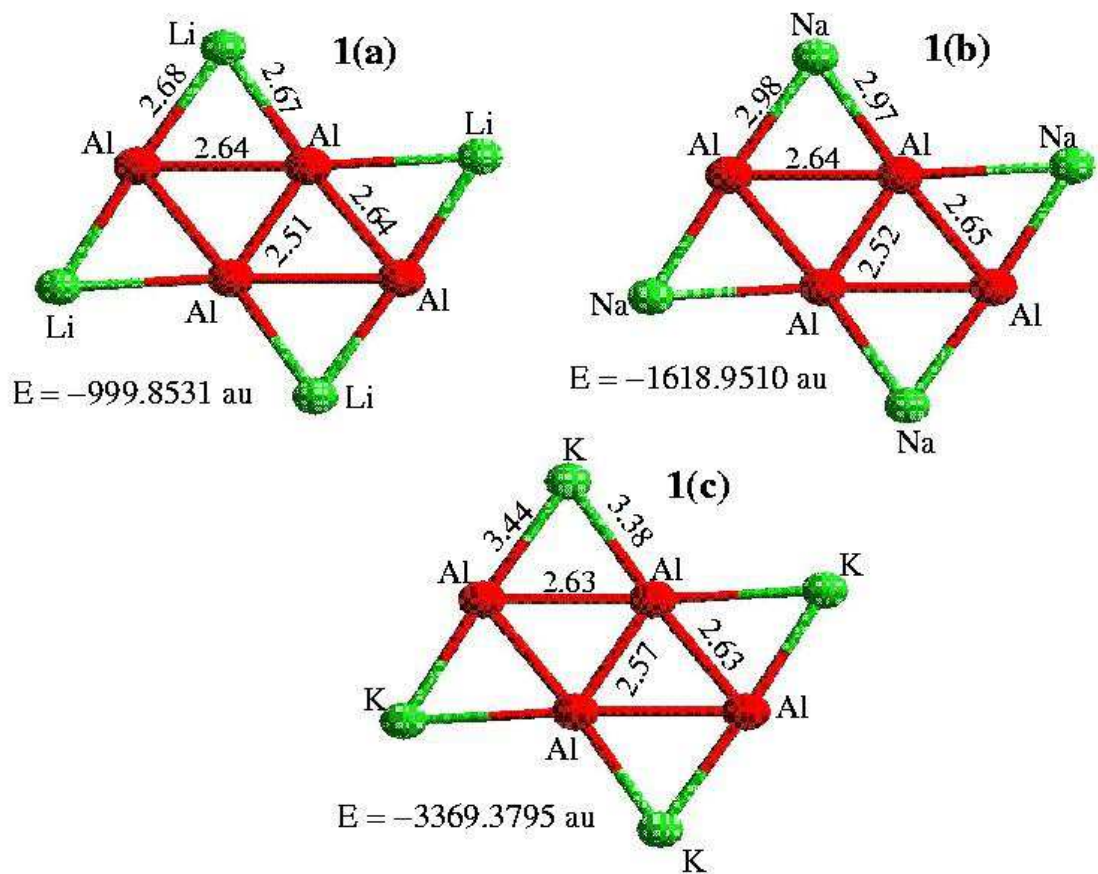}
\end{figure}

\newpage
\clearpage

\begin{figure}
\caption{Equilibrium ground state geometries for the
other set of Al$_{4}$Li$_{4}$,
Al$_{4}$Na$_{4}$ and Al$_{4}$K$_{4}$, very close
in energy to Figure 1. The footnote
of each structure contains the ground state energies in au.}
\includegraphics[scale=0.8] {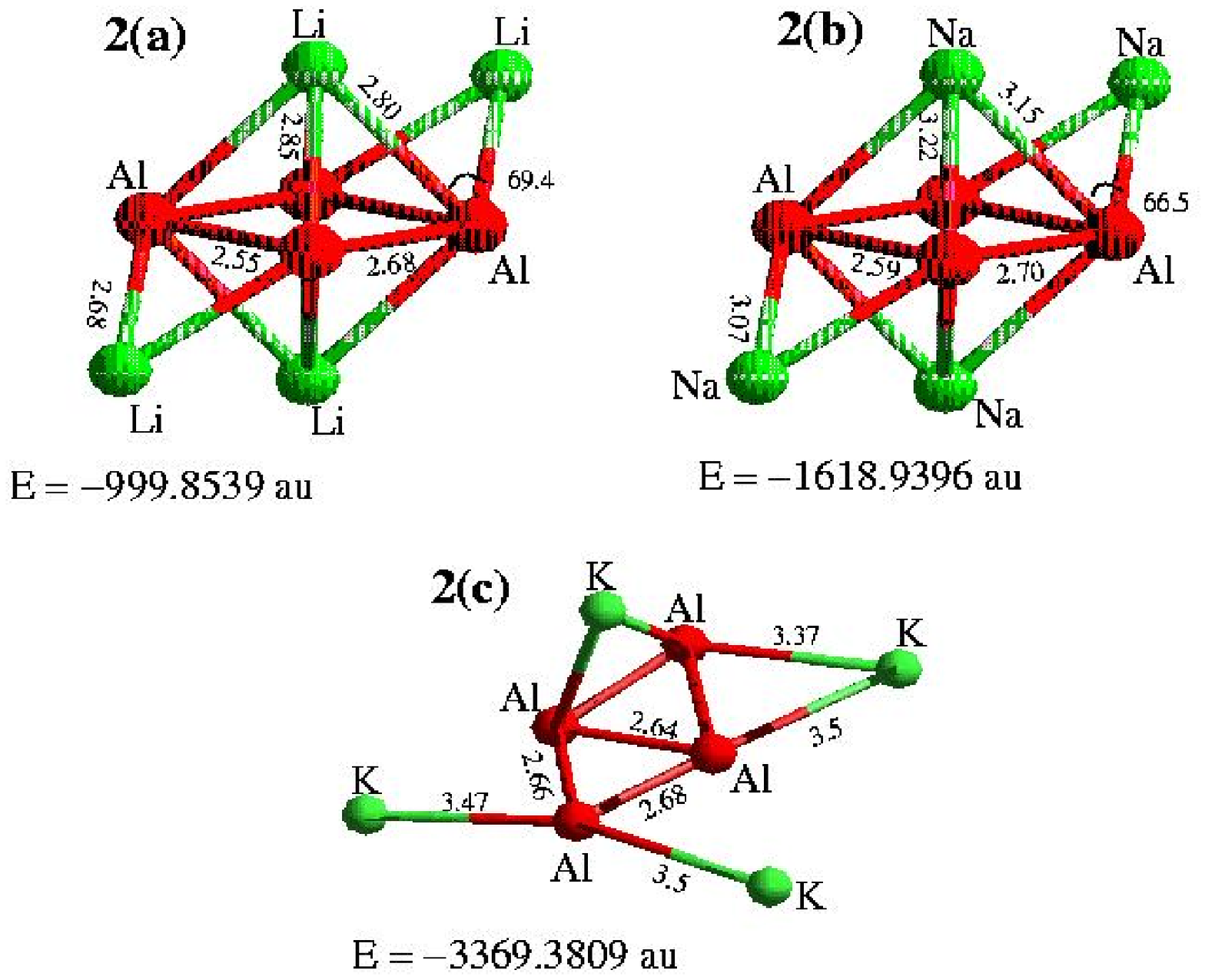}
\end{figure}


\begin{thebibliography}{99}

\bibitem{nlo1} (a) {\it Introduction to Nonlinear Optical Effects in Molecules
and Polymers}, P. N. Prasad and D. J. Williams (Wiley, New York, 1991);
 (b) {\it Nonlinear Optical Materials}, ACS Symposium Series 628, Eds. S.P.Karna and 
A.T.Yeates, Washington DC, 1996. 
\bibitem{wills} D. J. Williams, Angew. Chem. Int. Ed. Engl, 
{\bf 23}, 690, (1984).
\bibitem{marks} T. J. Marks and M. A. Ratner, Angew. Chem. Int. Ed. Engl,
{\bf 34}, 155, (1995).
\bibitem{pugh} (a) D. Li, T. J. Marks and M. A. Ratner, J. Phys. Chem, 
{\bf 96}, 4325, (1992). (b) S. Ramasesha and Z. G. Soos, Chem. Phys. Lett, {\bf 158}, 171, (1988).
\bibitem{Kuzyk} M. G. Kuzyk, Phys. Rev. Lett, {\bf 85}, 1218, (2000).
\bibitem{Helgaker} {\it Molecular Electronic-Structure Theory}, T. Helgaker, 
P. Jorgensen and J. Olsen (John Wiley \& Sons, New York, 2000) 
\bibitem{c1} B. K. Rao and P. Jena, J. Chem. Phys, {\bf 113}, 1508, (2000).
\bibitem{c2} (a) K. Wu, X. Chen, J. G. Snijders, R. Sa, C. Lin and
B. Zhuang, J. Cryst. Growth, {\bf 237}, 663, (2002). (b) G. Maroulis and C. Pouchan,
J. Phys. Chem.B, {\bf 107}, 10683, (2003).
\bibitem{sharan} S. Shetty, D. G. Kanhare and S. Pal, J. Phys. Chem. A, {\bf 108}, 628, (2004).
\bibitem{aroma} A. Kuznetsov, K. Birch, A. I. Boldyrev, X. Li, H. Zhai and
 L. Wang, Science, {\bf 300}, 622, (2003).
\bibitem{schleyer} Z. Chen, C. Corminboeuf, T. Heine, J. Bohmann and P. V. R. Schleyer, 
J. Am. Chem. Soc, {\bf 125}, 13930, (2003).
\bibitem{dft} (a). A. D. Becke, J. Chem. Phys. {\bf 98}, 1372, (1993); 
(b). C. Lee, W. Yang and R. G. Parr, Phys. Rev. B, {\bf 37}, 785, (1988).
\bibitem{gamess} M. W. Schmidt, K. K. Baldridge and J. A. Boatz et al. 
J. Comput. Chem, {\bf 14}, 1347, (1993).
\bibitem{kanhare} S. Chacko, M. Deshpande and D. G. Kanhare, Phys. Rev. B, 
{\bf 64}, 155409, (2001).
\bibitem{zerner} J. Ridley, M. C. Zerner, Theor. Chim. Acta
{\bf 32}, 111 (1973); A. D. Bacon, M. C. Zerner, Theor. Chim. Acta
{\bf 53}, 21 (1979).
\bibitem{mrdci1} R. J. Buenker and S. D. Peyerimhoff, Theor. Chim. Acta
{\bf 35}, 33 (1974).
\bibitem{mrdci2}  Z. Shuai, D. Beljonne and J. L. Bredas, J. Chem. Phys.
{\bf 97}, 1132 (1992).
\bibitem{mrdci3} D. Beljonne, Z. Shuai, J. Cornil, D. dos Santos and J. L.
Bredas, J. Chem. Phys. {\bf 111}, 2829 (1999).
\bibitem{sr1} S. Ramasesha and Z. G. Soos, Chem. Phys. Lett. {\bf 153},
171 (1988); Z. G. Soos and S. Ramasesha, J. Chem. Phys. {\bf 90}, 1067 (1989).
\bibitem{sr2} S. Ramasesha, Z. Shuai, J. L. Bredas, Chem. Phys. Lett.
{\bf 245}, 224 (1995); I. D. L. Albert and S. Ramasesha, J. Phys. Chem.
{\bf 94}, 6540 (1990); S. Ramasesha and I. D. L. Albert, Phys. Rev. B {\bf 42},
8587 (1990).
\bibitem{pati1} (a) A. Datta and S. K. Pati, J. Chem. Phys.,
{\bf 118}, 8420, (2003). (b) A. Datta and S. K. Pati, J. Phys. Chem. A, {\bf 108}, 
320, (2004). (c). D. Beljonne, J. Cornil, R. H. Friend, R. A. J. Janssen and J. L. Bredas,
J. Am. Chem. Soc. {\bf 118}, 6453, (1996). 
\bibitem{pati2} (a). S. K. Pati, S. Ramasesha, Z. Shuai and J. L. Bredas,
Phys. Rev. B {\bf 59}, 14827 (1999). (b). S. K. Pati, T. J. Marks and M. A. 
Ratner, J. Am. Chem. Soc.,{\bf 123}, 7287 (2001). 
\bibitem{pearson} R. G. Pearson, Inorg. Chem., {\bf 27}, 734 (1988).
\bibitem{nonaroma} {\it Advanced Organic Chemistry: Reactions, Mechanisms and Structure}, 
Jerry March, 4th edition, (John Wiley and Sons, 1992). 
\bibitem{tumbling} S. Ramasesha, Z. Shuai, J. L. Bredas, Chem. Phys. Lett.
{\bf 250}, 14 (1996).
\bibitem{rat} (a). J. A. Armstrong, N. Bloembergen, J. Ducuing,
and P. S. Pershan, Phys. Rev, {\bf 127}, 1918, (1962); (b). J. Ward, Rev. Mod. Phys. 
{\bf 37}, 1, (1965); (c). S. K. Pati, T. J. Marks 
and M. A. Ratner, J. Am. Chem. Soc, {\bf 123}, 7287, (2001).
\bibitem{marder} S. R. Marder, J. W. Perry, G. Bourhill, C. B. Gorman, B. G. Tiemann
 and K. Mansour, Science, {\bf 261}, 186, (1993).
\bibitem{expt} J. F. Ward and D. S. Elliot, J. Chem. Phys. {\bf 69}, 5438, (1978).
\bibitem{CLBO} J. Li, C-g, Duan, Z-q. Gu, D-s, Wang, Phys. Rev. B, {\bf 57}, 6925, (1998). 
\bibitem{papa1} M. Theologitis, G. C. Screttas, S. G. Rapis and M. G. Papadopoulos, 
Int. Jour. Quant. Chem. {\bf 72}, 177, (1999).
\bibitem{papa2} M. G. Papadopoulos, S. G. Raptis and I. N. Demetropoulos, Mol. Phys. 
{\bf 92}, 547, (1997).
\bibitem{karna} P. P. Korambath and S. P. Karna, J. Phys. Chem. A. {\bf 104}, 4801, (2000).
\bibitem{mercero} J. M. Mercero and J. M. Ugalde, J. Am. Chem. Soc. {\bf 126}, 3380, (2004).

\end{thebibliography}
\end{document}